\documentclass[sigconf,authorversion]{acmart}
\pdfoutput=1

\usepackage{listings}
\usepackage{aliascnt}

\newcommand{\pkcs}{\texttt{PKCS\#11}}
\newcommand{\users}{\ensuremath{\mathcal{U}}}
\newcommand{\usersSO}{\ensuremath{\mathcal{U}_{SO}}}
\newcommand{\usersKM}{\ensuremath{\mathcal{U}_{KM}}}
\newcommand{\attacked}{\ensuremath{\mathcal{A}}}
\newcommand{\candidatekeys}{\ensuremath{\mathcal{K}}}

\newcommand{\wwt}{\texttt{wrap\_with\_trusted}}
\newcommand{\trusted}{\texttt{trusted}}
\newcommand{\sensitive}{\texttt{sensitive}}
\newcommand{\wrap}{\texttt{wrap}}
\newcommand{\unwrap}{\texttt{unwrap}}
\newcommand{\decrypt}{\texttt{decrypt}}
\newcommand{\encrypt}{\texttt{encrypt}}
\newcommand{\extractable}{\texttt{extractable}}

\newcommand{\senc}[2]{\textsf{senc}(#1,#2)}
\newcommand{\sdec}[2]{\textsf{sdec}(#1,#2)}
\newcommand{\facts}[1]{\texttt{#1}}

\newcommand{\rewrulearray}[4]{ [~#2~]  & -\!\!\!-\!\!\![~#3~]\!\!\!\rightarrow & [~#4~] }

\AtBeginDocument{%
  \providecommand\BibTeX{{%
    \normalfont B\kern-0.5em{\scshape i\kern-0.25em b}\kern-0.8em\TeX}}}

\copyrightyear{2021} 
\acmYear{2021} 
\setcopyright{acmlicensed}
\acmConference[To appear at CCS'21]{To appear in proceedings of the 2021 ACM SIGSAC Conference on Computer and Communications Security}{November 15--19, 2021}{Virtual Event, Republic of Korea}
\acmBooktitle{Proceedings of the 2021 ACM SIGSAC Conference on Computer and Communications Security (CCS '21), November 15--19, 2021, Virtual Event, Republic of Korea}
\acmPrice{15.00}
\acmDOI{10.1145/3460120.3484785}
\acmISBN{978-1-4503-8454-4/21/11}
\begin{document}

\title{A Formally Verified Configuration \\for Hardware Security Modules in the Cloud}


\author{Riccardo Focardi}
\email{focardi@unive.it}
\affiliation{%
  \institution{DAIS, Ca' Foscari University}
  \streetaddress{via Torino 155}
  \city{Venice}
  \country{Italy}
  \postcode{30172}
}

\author{Flaminia L. Luccio}
\email{luccio@unive.it}
\affiliation{%
  \institution{DAIS, Ca' Foscari University}
  \streetaddress{via Torino 155}
  \city{Venice}
  \country{Italy}
  \postcode{30172}
}

\renewcommand{\shorttitle}{A Formally Verified Configuration for Hardware Security Modules in the Cloud}

%%% Theorems and autoref
\theoremstyle{acmdefinition}
\newtheorem{guideline}{Rule}
\newtheorem{fact}{Fact}
\newtheorem{claim}{Claim}
\newaliascnt{def}{theorem}
\newtheorem{mydef}[def]{Definition}
\aliascntresetthe{def}
\newcommand{\guidelineautorefname}{Rule}
\newcommand{\claimautorefname}{Claim}
\newcommand{\factautorefname}{Fact}
\renewcommand{\sectionautorefname}{Section}
\renewcommand{\subsectionautorefname}{Section}
\newcommand{\defautorefname}{Definition}

\begin{abstract}
Hardware Security Modules (HSMs) are trusted machines that perform sensitive operations in critical ecosystems. They are usually required by law in financial and government digital services. The most important feature of an HSM is its ability to store sensitive credentials and cryptographic keys inside a tamper-resistant hardware, so that every operation is done internally through a suitable API, and such sensitive data are never exposed outside the device. HSMs are now conveniently provided in the cloud, meaning that the physical machines are remotely  hosted by some provider and customers can access them through a standard API. The property of keeping sensitive data inside the device is even more important in this setting as a vulnerable application might expose the full API to an attacker. Unfortunately, in the last 20+ years a multitude of practical API-level attacks have been found and proved feasible in real devices. The latest version of \pkcs{}, the most popular standard API for HSMs, does not address these issues leaving all the flaws possible. In this paper, we propose the first secure HSM configuration that does not require any restriction or modification of the \pkcs{} API and is suitable to cloud HSM solutions, where compliance to the standard API is of paramount importance. The configuration relies on a careful separation of roles among the different HSM users so that known API flaws are not exploitable by any attacker taking control of the application. We prove the  correctness of the configuration by providing a formal model in the state-of-the-art Tamarin prover and we show how to implement the configuration in a real cloud HSM solution.
\end{abstract}




\maketitle

\section{Introduction}
%!TEX root = main.tex

The increasingly massive digitization we have been witnessing in recent years is leading to a pervasive use of cryptography for daily tasks. This phenomenon is even more pronounced in critical ecosystems such as financial and governmental ones. In these contexts it is often mandatory to use special machines called Hardware Security Modules (HSMs) that allow applications to perform critical operations internally, without exposing the cryptographic keys. In fact,  the main feature of an HSM is to keep sensitive data inside a tamper-resistant hardware so that in the event of a physical attack it is not possible to extract the cryptographic keys and produce an identical copy of the HSM for malicious purposes. In other words, resistance to intrusion makes an HSM unique, similarly to a smartcard: an attacker who wants to carry out illicit cryptographic operations should necessarily have access to the device since its cloning, which requires a copy of the secret data kept in the device, should not be possible.

This special hardware is very expensive and is not tailored to a specific application. Its standard API, named \pkcs{}, is in fact a general purpose API for cryptographic operations with some peculiar features. Keys have attributes that determine their role and usage so, for example, the value of a \sensitive{} key cannot be directly accessed from outside the device. However, keys can be exported encrypted under so called \wrap{} keys, in order to be securely stored off the device or imported and shared with other HSMs. Unfortunately, in the last 20+ years a number of API level attacks on \pkcs{} have appeared in the literature. In his MSc thesis, Clulow \cite{clulow03thesis} showed that an astonishingly trivial attack was possible: given a target sensitive key, it was possible to generate another key with both \wrap{} and \decrypt{} attributes set. This allowed for a trivial \emph{wrap-then-decrypt} attack: the sensitive key was first wrapped with the \wrap{} key, producing a ciphertext outside the device which was then sent back to the device for decryption under the very same key. The coexistence of these conflicting roles for the same key made it possible to obtain the plaintext value of a sensitive key off the device after two simple API calls: first a \emph{wrap} and then a \emph{decrypt}. 

After Clulow's seminal work, many researchers have pointed out more attack variants and investigated possible fixes to the API. A practical implementation of a secure wrapping mechanism have been even suggested to Oasis, the current maintainer of the \pkcs{}, with no success \cite{GrahamProposal}: none of the proposed fixes have received enough attention, and the latest version of the \pkcs{} standard \cite{PKCS11v3} does not contain any guideline on how to prevent this attacks in practice and, in fact, does not even mention them. On the contrary, referring in a broad sense to API-level attacks, the latest version of \pkcs{} usage guide \cite[Section 3.1]{PKCS11usage}, which dates back to November 2014, surprisingly states that: ``We note that none of the attacks just described can compromise keys marked sensitive, since a key that is sensitive will always remain sensitive''. While this is certainly true when it comes to modifying attributes, the attack described above allows \sensitive{} keys to be leaked in the clear, breaking the required security property. Hence, we are faced with a disturbing situation where expensive hardware enforced by law in security-critical environments is subject to attacks that are not even mentioned in the standard API documentation.

The only available defense was introduced in version 2.40 of the \pkcs{} standard \cite{PKCS11v2.40}: a special attribute \wwt{} that, when set on \sensitive{} keys, only allows wrapping under keys marked as \trusted{} by a privileged HSM user called Security Officer. However, this attribute is not mandatory for sensitive keys and it is unclear how these special trusted keys should be generated and managed. As previously discussed, the documentation does not provide guidance on this important attribute, and does not mention the fact that not using it exposes the HSM to well-known vulnerabilities. In fact, not using \wwt{} completely voids the adoption of an HSM as any attacker gaining access to the API could leak any key with \wwt{} attribute unset.

Particular installations might use ad-hoc solutions to prevent the above mentioned vulnerabilities, such as disabling key wrapping and any other dangerous key management functionalities in production devices, or replace them with nonstandard proprietary APIs as, e.g., \cite{BCFS-ccs10,KunCSF19,centenaro09type,KunPOST15,Oak2017}. However, these solutions break portability and, in general, it would be preferable to use the full API functionality in a secure way rather then cutting it down to a secure subset, or modifying it in nonstandard ways. Moreover, proprietary solutions are not publicly documented and cannot be validated by the scientific community. In many cases, this \emph{security-by-obscurity} approach only provides a false sense of security and many proprietary solutions exhibit vulnerabilities once they have been reverse-engineered or just leaked to the public. 
The numerous attacks on cryptographic protocols for automotive systems are a representative example (e.g., \cite{GarciaMegamos,GarciaKeyless}). 

In recent years, we have witnessed the rapidly growing phenomenon of cloud HSMs. Several cloud service providers give the possibility to use real HSM clusters managed entirely in the cloud and accessible via remote APIs \cite{AWSHSM,GoogleHSM,IBMHSM,UtimacoHSM}. This phenomenon radically changes the attacker model. With classic, physically reachable HSMs, it was possible to use specific procedures for managing keys, which could only be performed by users with physical access to the machines; with cloud HSMs these procedures must necessarily take place remotely via the API offered by the service provider. Furthermore, with classic HSMs it was possible to customize their configuration based on the particular application, for example by disabling some critical functionalities, and possibly using proprietary APIs;
with cloud HSMs this possibility no longer exists as they only provide standard APIs in order to work with any \pkcs{}-based application. In fact, the advent of cloud HSMs  makes the weakness of their APIs even more critical and requires an accurate and correct use of \trusted{} keys. In addition, it becomes extremely important to apply the least privilege principle to applications to prevent a vulnerability from allowing the attacker to access the (vulnerable) key management functions, giving the ability to extract sensitive keys as cleartext. Unfortunately, as already discussed, the API documentation does not contain any guidelines in this regard.

\medskip
In this paper, we propose the first configuration for HSMs based on a careful separation of roles so that the critical key management operations are only performed by special users in a way that production applications cannot exploit the (flawed) API to extract keys. As we discussed earlier, this is very important, especially in a cloud setting, where the HSM is operated remotely and a vulnerability in one application might compromise the security of the whole device, leaking all of the sensitive keys. Interestingly, our solution does not require any limitation of the API for the applications: any HSM user has access to the full API, however only the Security Officer can mark a key as \trusted{} as defined in the standard. We define the following categories of users:

\begin{description}

\item[Normal Users (NU)] These users are accessible by production applications, so it is of paramount importance that: $(i)$ they can use the full API, in order to maximize compatibility; $(ii)$ they cannot exploit API-level attacks to leak sensitive keys, as they are the most exposed HSM users: in fact, an attacker might exploit a vulnerability in production application to gain access to the HSM API. Intuitively, the keys of these users that need to be wrapped under other keys will always be marked as \wwt{}.

\item[Key Manager (KM)] These users perform key management operations using the standard API in a controlled way. They are responsible for handling the \trusted{} keys that will be used to wrap other sensitive keys. The life-cycle of these keys is very delicate. For example, they should never be wrapped under other keys and never be allowed to encrypt or decrypt data. The KM accounts should only be accessed by special management applications (or directly by humans), and their credentials should never be used in production applications.

\item[Security Officer (SO)] The Security Officer is a special user which adheres to the \pkcs{} standard and has no access to the full API \cite[Section 2.4]{PKCS11usage}. As such, the SO mainly performs administrative tasks, such as creating other HSM users and marking keys as \trusted{}. Specifically, we require the SO to mark as \trusted{} only the keys generated by a Key Manager. As for KMs, the SO account should only be accessed by special management applications (or directly by humans), and its credentials should never be used by production applications.

\end{description}  
Intuitively, if we assume that the management of \trusted{} keys is done by special users (KMs and SO) and is not accessible by production applications, \wwt{} keys will never be leaked in the clear, even using all known API-level attacks. The tricky part of the configuration is to define precisely what the KMs and SO should or should not do. For example, enabling decryption on a \trusted{} wrapping key would allow the trivial Clulow's wrap-then-decrypt attack previously discussed. Encryption is also dangerous as an attacker could encrypt a known key and then import it to the device using unwrap: these malicious keys should never be marked as \trusted{}. Finally, if a \trusted{} key can be wrapped under another trusted key then an attacker might \emph{reimport} it in the device with different attributes enabling the previous attacks \cite{DKS-jcs09}. Our separation of roles is based on the crucial assumption that keys generated by one user can  be made available to other HSM users but key management operations, such as attribute change, will only be permitted to the key owner. We will show that this form of mild access control is implemented in real cloud HSM solutions.

Given the well-documented subtleties behind this unfortunate API, it is of paramount importance to provide evidence of the correctness of the proposed configuration. We decided to  model it using the state-of-the-art Tamarin prover, a powerful tool that allows for  proving security properties of cryptographic protocols. In our model, the attacker has full access to the API but cannot impersonate neither the Security Officer nor the Key Manager. In this way, we cover any vulnerable application with access to the HSM APIs which cannot tamper with the management of \trusted{} keys.
The model represents a significant subset of \pkcs{} including symmetric key cryptography, key management, core key attributes and users, and we prove our configuration correct under various assumptions on the behavior of the KMs and  SO. Our policy is simple to implement and we show how to apply it to the AWS CloudHSM solution \cite{AWSHSM} which supports the required form of access control between users and keys. The simplicity of the policy comes from the simplicity of the \wwt{} mechanism that, unfortunately, does not offer enough flexibility to implement more sophisticated secure configurations.

\subsubsection*{Contributions} The main contributions of this work are summarized below:
\begin{enumerate}

\item We explore for the first time the usage of users' roles for securing \pkcs{}. In particular, we introduce special users called Key Managers that are in charge of creating and managing the \trusted{} keys. The keys generated by the KMs are made available to other HSM users but key management will only be permitted to the key owner;

\item Based on the users' roles, we propose the first practical HSM configuration that does not require any restriction or modification of the \pkcs{} API and is suitable to cloud HSM solutions. The configuration comprises a set of rules for the secure usage of \trusted{} and \wwt{}  attributes and we claim its correctness through an informal analysis based on existing API-level attacks;

\item We show how the rules can be directly applied to the AWS CloudHSM which
provides a suitable access control over users and keys. Interestingly, AWS CloudHSM allows for separating key usage from key management, which is one of the crucial assumption of our configuration;

\item Finally, we provide a formal model of a significant core of \pkcs{} and we use it to model the proposed secure configuration rules as constrains on the behavior of KMs and SO. We use the Tamarin tool to  automatically verify that sensitive keys are never leaked to normal users, even when they are in full control of the API, assuming that KMs and SO behave correctly. The proof is fully automated and the model can be easily adapted to check variants of the proposed configuration.

\end{enumerate}

\subsubsection*{Related work}
\label{sec:related}
%!TEX root = main.tex

The first security vulnerability that may properly be called an API-level
attack was discovered by Longley and Rigby in the early 1990s
\cite{longley92automatic}. 
Although the device was not identified at the time, it was later discovered that it was an HSM manufactured by Eracom and used in the ATM network.

In 2001, Anderson published an attack on key loading procedures on another similar module manufactured by Visa \cite{anderson00correctness}, and the term \emph{security API} was coined by Bond and Anderson \cite{bond01attacks,bond01api} in two subsequent papers presenting other attacks. Clayton and Bond showed how computationally intensive attacks could be implemented against a real IBM device using programmable FPGA hardware \cite{clayton02experience}. Independently from the Cambridge group, an MSc thesis by Clulow gave more examples of attacks, mostly specific to the PIN translation and verification commands offered by the API of Prism HSMs \cite{clulow03thesis}. Clulow also published the first attacks on the industry standard for cryptographic key management APIs, \pkcs{} \cite{clulow03pkcs11}.

This proliferation of vulnerabilities has stimulated research results regarding the analysis of security APIs that has helped understanding in depth the nature of key management attacks. A first effort to apply general analysis tools appeared in \cite{youn05robbing}, but the researchers were unable to discover any new attacks and could not conclude anything about the security of the device. The first automated analysis of \pkcs{}  with a formal statement of the underlying assumptions was presented in \cite{DKS-jcs09}. When no attacks were found, the authors were able to derive precise security properties of the device. In \cite{BCFS-ccs10}, the model was generalized and provided with a reverse-engineering tool that automatically refined the model depending on the actual behaviour of the device. When new attacks were found, they were tested directly on the device to get rid of possible spurious attacks determined by the model abstraction. The automated tool of \cite{BCFS-ccs10} successfully found attacks that leaked the value of sensitive keys on real devices. In \cite{KunCSF19}, \emph{authenticated wrapping} was analyzed and proved correct under some assumptions. The idea was that the attributes were wrapped together with the key value so that they were preserved when the keys were exported and reimported. This mechanism would provide a fundamental improvement in \pkcs{} security but, unfortunately, it has not yet been included in the standard. Other previous attempts to request its inclusion have so far failed (see, e.g., \cite{GrahamProposal}).

The works discussed so far do not take into account the attribute \wwt{}  which was introduced in version 2.20 of \pkcs{}. The first analysis of the \wwt{} attribute appeared in \cite{FroschleS11}, where the authors showed that when \trusted{} keys have a fixed, pre-determined set of attributes, all keys with \wwt{} set are secure. Intuitively, the work provided formal evidence that the basic mechanism was secure, when used in a very controlled and limited way. In \cite{CentenaroFL13} and \cite{KunPOST15} the analysis was generalized to other key configurations and was proved correct via typing and automated verification, respectively. In \cite{Oak2017} a \pkcs{} configuration proposed in \cite{BCFS-ccs10} was proved secure on a computational model of \pkcs{}, using \wwt{} to prevent key cycles which, by the way, is not the primary purpose of this attribute. All of these works assume that the attributes of keys are immutable. While this property is certainly desirable for a cryptographic API, \pkcs{} allows for attribute changes and removing this functionality might break a multitude of applications. Therefore, the applicability of these proposals is confined to specific applications and is rather limited for what concerns cloud HSMs which aim to offer a complete \pkcs{} interface.

Our work is also based on the \wwt{} attribute, but our proposal requires minimal constraints on key management by leveraging user roles. To the best of our knowledge, our configuration is the first one that does not require 
%\pkcs{} hacks or feature reductions: 
any modification or reduction of \pkcs{}: it simply requires some particular users to be trustworthy and behave according to strict guidelines, making it directly applicable to real cases.  In particular, we do not impose any restriction on production users, who can use the full API including the change attribute functionality, and we only limit what Key Managers and the Security Officer can do. 

The Tamarin prover \cite{MSCB13} has been successfully used to prove the correctness of many real world protocols (see, e.g., \cite{TamarinCCS20,TamarinUSENIX20,TamarinUSENIX20b,TamarinNDSS19,CSFpuf2020}). Following this line of work, we model a core subset of \pkcs{} in Tamarin obtaining a fully mechanized proof of our secure configuration for an unbounded number of sessions, users and keys. This allows for easily validating possible model extensions and variants. The automated analyses of \cite{KunPOST15} was also performed using Tamarin as a backend, but the model was specified in a process calculus which was translated into a Tamarin model so, to the best of our knowledge, our model of the \wwt{} mechanism is the first one that is directly expressed in the Tamarin prover language. A peculiar  feature that makes the analysis of security APIs hard is that they have a non-monotonic mutable global state \cite{DKS-jcs09}. The assumption on immutable attributes of \cite{KunPOST15} simplifies this aspect. Our model, instead, supports non-monotonic attributes that we made treatable through new, suitable simplifications (cf. \autoref{sec:model:simplification}). Moreover, we model for the first time access control to keys.
% Moreover, as we discussed earlier and differently from our approach, paper \cite{KunPOST15} assumes that key attributes are immutable. 
In \cite{KunCSF19} a model of authenticated wrapping was expressed and verified in Tamarin. The proposed authenticated wrapping mechanism is based on unique counters, a key hierarchy and authenticated handles, which are not in the \pkcs{} standard. In fact, as we discussed earlier, the aim of the paper was to propose new, provable secure extensions of the \pkcs{} standard and not, as we do here, to analyze the security of \pkcs{} configurations.

\subsubsection*{Paper organization}

The paper is organized as follows: in \autoref{sec:background} we recall some notions related to the \pkcs{} API and to the Tamarin prover. In \autoref{sec:configuration} we present our new HSM configuration and we apply it to the AWS CloudHSM solution. In  \autoref{sec:formal}  we formalize our configuration in Tamarin and we automatically prove its security. In  \autoref{sec:conclusion} we provide some concluding remarks.

\label{sec:introduction}

\section{Background}
\label{sec:background}
%!TEX root = main.tex
In this section, we first introduce the \pkcs{} standard and API-level attacks (\autoref{sec:pkcs}), then we briefly illustrate the Tamarin prover (\autoref{sec:tamarin}) which will later be used to prove the security properties of the proposed HSM configuration.

\subsection{The \pkcs{} Standard}
\label{sec:pkcs}

The RSA Public Key Cryptography Standard \#11 (\pkcs{}), first proposed  in 1995 as Version 1.0, describes an API named `Cryptoki' for cryptographic hardware such as HSMs, cryptographic tokens and smart cards. It supports both cryptographic and key management functions with the general idea that all operations are performed inside the device so that cryptographic keys are never exposed in the clear to external applications. The latest version of the standard, taken over by OASIS in 2012,  is Version 3.0 \cite{PKCS11v3}.

The \pkcs{} API is used by applications as follows. The application starts a \textit{session} with the device identifying either as a Security Officer (SO), or as a normal user. In this phase, there is no defense against a rogue host or application, as the user PIN may be intercepted. However, this should not allow an attacker to compromise a sensitive key ``since a key that is sensitive will always remain sensitive'', as stated in the latest version of \pkcs{} usage guide \cite[Section 3.1]{PKCS11usage}, which dates back to November 2014. After a session is established, the application  may access the \textit{objects}, such as keys and certificates, stored in the device. Objects are accessed through \textit{handles} that point to them and do not disclose  any information about the object they point to. New objects, pointed by fresh handles, are either created with a key generation command or by unwrapping an encrypted blob, as described below.

Objects have \textit{attributes} that specify properties or roles. Many of them are boolean flags that can be set or unset. For example, attribute \sensitive{}, when set,  protects a key from being read in the clear, and cannot be unset, to prevent trivial attacks. A key can be exported from the device encrypted under another key only if its attribute \extractable{} is set, and once unset it cannot be set again. The attributes \encrypt{} and \decrypt{} indicate keys that can be respectively used to  encrypt and decrypt data. The attribute \wrap{}, instead, is used to encrypt another key that has the \extractable{} attribute set and then to export it as a ciphertext. Similarly, \unwrap{}  is used to import keys: it decrypts an encrypted key and imports it in the device as a new object, returning a corresponding fresh handle. Thus, wrap and unwrap functions allow for exporting and reimporting (\extractable{}) keys encrypted under so called \emph{wrapping} keys.  All the cryptographic functions use key handles to refer to keys so that their value is never exposed outside the device. For example, wrap takes as input two handles that refer respectively to the key to be wrapped and the wrapping key, while encrypt takes as input data bytes and one handle that refers to the encryption key.

\begin{figure}[bt]

\begin{center}
\fbox{

\begin{minipage}[c]{3in}
Handle $a_1$, associated to key $k_1$, has attributes \sensitive{} and \extractable{} set and handle $a_2$, associated to key $k_2$, has attributes \wrap{} and \decrypt{} set; $c$ denotes the  encryption of $k_1$ under key $k_2$ using a symmetric key cipher.
The attack leaks the sensitive key $k_1$ referenced by $a_1$ in the clear.  

\medskip
\medskip

\begin{centering}
\begin{tabular}{rcl}
\textsf{\textbf{Wrap}($a_1$,$a_2$)}  & $\rightarrow$ & $c$  \\
\textsf{\textbf{Decrypt}($c$,$a_2$)} & $\rightarrow$ & $k_1$\\
\end{tabular}

\end{centering}
\medskip

\end{minipage}
}
\end{center}

\caption{\pkcs{} wrap-then-decrypt attack.}
% \cite{clulow03PKCS11}}
\label{fig1}

\end{figure}

\paragraph{API-level attacks on \pkcs{}} The first attacks to the \pkcs{} API were introduced by Clulow in 2003  \cite{clulow03pkcs11}. The simplest one is called  \textit{wrap-then-decrypt}: the attacker first wraps a \sensitive{} key, and then decrypts it. More precisely, the attacker has two handles: $a_1$, associated to a key $k_1$, with attributes  \sensitive{} and \extractable{} set and $a_2$, associated to a key $k_2$, with attributes \wrap{} and \decrypt{} set. The attacker is able to leak the sensitive key $k_1$ in the clear by first wrapping it under $k_2$ obtaining ciphertext $c$, and then decrypting it using key $k_2$. Note that, wrap is possible because of the \wrap{} attribute set on $a_2$ and the \extractable{} attribute set on $a_1$, while decrypt is possible because of the \decrypt{} attribute set on $a_2$. When the device executes the operations, it cannot distinguish between keys and plaintexts and the decrypted key is given as output to the attacker. The attack is illustrated in Figure~\ref{fig1}.

There exists a \emph{dual} version of this attack called \emph{encrypt-then-unwrap} in which handle $a_2$ has attributes \unwrap{} and \encrypt{} set. The attacker encrypts a known key $k_A$ under $a_2$ and then unwraps it, importing it in the device with fresh handle $a_3$ and attribute \wrap{} set. Note that, when a key is unwrapped it is possible to specify its attributes. The attacker can now wrap $a_1$, which refers to the sensitive key $k_1$, under $a_3$ obtaining the encryption of $k_1$under the known key $k_A$. Decryption can be then performed by the attacker independently of the device. 

In principle, these attacks could be prevented by forbidding conflicting roles on keys such as \wrap{}, \decrypt{} and \unwrap{}, \encrypt{}. However attributes can be set and unset liberally so, for example, \wrap{} could be unset before \decrypt{} is set and, even if these conflicts could be prevented on a single handle in one device, subtler attacks exist that exploit the presence of a key under multiple handles, possibly on different devices. For example, a wrapped key could be imported twice providing two copies with conflicting roles or, equivalently, an existing key could be wrapped and then unwrapped to provide a new instance of the key with a conflicting role. Note that, these latter attacks could be carried out on separate devices sharing a wrapping key, making it very difficult to track conflicting attributes on the same key: the sensitive key could be wrapped under one copy of the key on the first device and decrypted under another copy of the key on the second device. The interested reader can refer to \cite{DKS-jcs09} for a comprehensive analysis of all attack variants and to \cite{FocardiLS10} for a more detailed introduction to API-level attacks.

\subsubsection*{Trusted wrapping keys} The only existing mechanism in the standard that can prevent API-level attacks is the \wwt{} attribute, introduced in Version  2.20, that forces a key to be  wrapped only under keys that have attribute \trusted{} set. Once set, the \wwt{} attribute cannot be unset and the only user that can mark a key as \trusted{} is the Security Officer. Note that, while this attribute can potentially prevent previous attacks, in the standard it is not mandatory to set \wwt{} on sensitive keys and there is no mention of how \trusted{} keys should be generated and managed, vanishing all  advantages of this mechanism. Some works proved that an accurate use of this attribute in a very controlled an limited way may work correctly, but they put limits to its application and assume that the attributes of keys are immutable, which is not realistic  as it reduces the API functionality and breaks compliance with the \pkcs{} standard \cite{FroschleS11,CentenaroFL13,KunPOST15,Oak2017,AdaoFL13}.
 
\subsection{The Tamarin Prover}
\label{sec:tamarin}

The Tamarin prover  is a very powerful verification tool \cite{MSCB13} that has been successfully used to prove the correctness of many real world protocols (see, e.g., \cite{TamarinCCS20,TamarinUSENIX20,TamarinUSENIX20b,TamarinNDSS19,CSFpuf2020}). Attackers and  protocols are specified using multiset rewriting rules, and properties as first-order logic formulas that admit quantification over timepoints and  messages.
Proofs can either be done interactively, by combining  automated proof search with manual guidance, or in a fully automated way using heuristic searches. In this second scenario, if Tamarin  terminates, it either provides a proof of correctness of all system traces, or an attack as a counter-example trace of the system.

\subsubsection*{Terms, facts and special facts} In Tamarin, \emph{terms} are used  to symbolically represent data and messages and functions, e.g.,  $\senc{\_}{\_}$,  $\sdec{\_}{\_}$ are used to express symmetric key encryption and decryption. Tamarin supports equational theories, e.g., $\sdec{\senc{m}{k}}{k}$ $ =_E$ $ m$ that makes it possible to deconstruct $\senc{m}{k}$ and recover $m$ from the term $\sdec{\senc{m}{k}}{k}$; \emph{facts} $\facts{F}(t_1,\ldots,t_n)$ are used to store state information  while \emph{special facts} have a particular semantics. For example, $\facts{In}(m)$, $\facts{Out}(m)$ are used for receiving and sending message $m$ and $\facts{Fr}(n)$ to generate a fresh name $n$, not known by the attacker.

\subsubsection*{Rules, execution and properties} The systems starts from an  {\em initial state} , i.e., the empty multiset, and transitions are specified by a {\em set of rules}, which are in the form: 
\begin{eqnarray*}
\rewrulearray{}{ p_1, \ldots, p_m }{ a_1, \ldots, a_k }{ c_1, \ldots, c_n} 
\end{eqnarray*}
where $p_1, \ldots, p_m$ are the premise facts that must belong to the state to enable the rule. Variables inside facts are instantiated once they match the facts in the state. When the rule is fired different actions are taken: the premise facts are removed from the state, unless they are specified as persistent using the symbol $!$; action facts $a_1, \ldots, a_k$ become part of the execution trace and can  be used to express properties of the protocol using first-order logic formulas with timepoints; finally, conclusion facts $c_1, \ldots, c_n$ are added to the state. Variables of  action and conclusion  facts are bound to the premise facts variables and are instantiated accordingly. An alternating   sequence  of   states   and   rule   instances defines a protocol execution. Note that, there are built-in rules to model a standard Dolev-Yao attacker \cite{dolev83security}.

The following first-order formula is a simple example of a property that can be expressed in Tamarin:
\[
\begin{array}{l}
\exists \ \facts{U}, \facts{h}, \facts{\#i} ~.~ \facts{SetAttr(U,h,a)}@\facts{i} 
\end{array}
\] 
It states that action fact $\facts{SetAttr(U,h,a)}$ has occurred at some time \facts{i}. The symbol \facts{\#} in front of \facts{i} is used in the quantifiers to indicate that variable \facts{i} represents a timepoint. We will use this action fact in \autoref{sec:model:keys} to model a user \facts{U} setting attribute \facts{a} on handle \facts{h}.

\section{A secure HSM configuration}
\label{sec:configuration}
%!TEX root = main.tex
As we have pointed out in the previous sections, 
the \pkcs{} API is flawed and there is no easy way to fix it without reducing its functionality or resorting to non-standard extensions. In the extensive literature on the subject, it has been pointed out many times that the excessive liberality of the API, especially in terms of changing key attributes, is the main source of the problem. Nevertheless, \pkcs{} is still the standard API for cryptographic hardware and HSMs in particular, and it is of paramount importance to investigate general guidelines to use it in a proper way, limiting the attack surface as much as possible. Using a flawed API to access a very expensive  secure hardware is quite disturbing, and we believe that the growing adoption of cloud HSM solutions will eventually provide a huge attack surface, completely nullifying the promised enhanced security. We also have to keep in mind that the use of HSM is required by law in many settings, and discovering that the security of these devices is poor, after so many years of warnings from the academic community, would have very negative consequences.

In this section, we make a new proposal for a secure HSM configuration which does not require any change in the \pkcs{} API. As a consequence, our proposal is fully compliant with the standard API and can be adopted immediately. The configuration relies on few requirements for user management in HSMs that we believe might be easily incorporated in any cloud HSM solution. Interestingly, we found that AWS CloudHSM already provides the necessary ingredients to implement our configuration.

The section is organized as follows: in \autoref{sec:attacker} we present the attacker model and in \autoref{sec:security} we give a definition of HSM security; in \autoref{sec:users} we describe the user roles and, based on them, in \autoref{sec:rules} we define our secure configuration; in \autoref{sec:sec_analysis} we show how known attacks are prevented by the secure configuration and claim its security in a general case; finally, in \autoref{sec:implementation} we show how to implement our configuration in a real cloud HSM solution.

\subsection{Attacker Model}
\label{sec:attacker}
Our attacker model is based on the classic API-level security approach where the attacker is assumed to gain access to the device's API, targeting its sensitive keys.
Notice that, once the API is available to the attacker, any cryptographic operation will be granted. For example, the attacker will be able to decrypt any ciphertext encrypted under the HSM symmetric keys and to produce any digital signature using HSM private keys. Therefore, API access naturally gives the attacker the power to \emph{use} sensitive keys in arbitrary ways. However, the advantage of HSMs is that the attacker should not be able to extract sensitive keys and \emph{clone} the HSM functionality so to have unbound access to those keys. 

The difference might seem insignificant but in fact is very important and, by the way, is the only extra layer of protection provided by this secure cryptographic hardware. The only place where sensitive keys should be used is inside the HSM: without access to the device it should be impossible to use its keys. A typical use case is an HSM used by a Certification Authority (CA) to sign public key certificates. An attacker gaining access to the HSM might create rogue certificates but should not be able to extract the root private key from the HSM. A temporary access to the HSM due to some vulnerability could be fixed by revoking the generated rogue certificates while leaking the root private key would require to revoke it, invalidating any existing certificate signed by that CA. 

Moreover, continuous access to the actual HSM would be detectable in the medium/long term while having a copy of the HSM keys would make the attack persistent and definitively harder to spot, as the attacker could locally use the keys  without any access to the HSM. By analogy, an HSM could be thought of as a very powerful smartcard: physically accessing the smartcard gives temporary privileges to an attacker, e.g. accessing a building, but cloning it would provide unlimited access.

\subsubsection*{Refining the attacker model} Since the API is flawed, full API access allows for extracting sensitive keys and makes any attempt of defining a secure configuration void. In other words, the general API-level attacker model is too strong and requires changing the API in order to achieve some security (see, e.g., \cite{BCFS-ccs10,KunCSF19,centenaro09type,KunPOST15,Oak2017,AdaoFL13}). Our strategy is to exploit a form of mild access control over keys that makes it possible to define different HSM users with specialized roles. In this way, we can effectively isolate critical key management operations that require some trust. In particular, we assume that keys are shared among users, and we let the owner of the key do whatever operation with it, while we forbid other users to change the attributes of the keys they do not own:

\begin{mydef}[Key owner]
\label{def:owner}
An HSM user creating a key is its \emph{owner} and can perform any operation over that key. Other users can use keys that they do not own but they cannot modify their attributes.
\end{mydef}
We refine accordingly the attacker model by assuming that the attacker only impersonates a subset of the HSM users. This makes sense in practice as accessing the HSM from production applications will be more vulnerable to attacks than accessing the HSM from management applications used by administrators. We let \users{} denote all the HSM users:

\begin{mydef}[Attacker model]
\label{def:attacker}
Given a subset of the HSM users $\attacked{} \subseteq \users{}$, we say that an HSM is under  \attacked{}-attack if the attacker can access the HSM API as one of the users in \attacked{}.
\end{mydef}

Intuitively, we will investigate HSM security assuming that a portion of the users, namely $\users{} \setminus \attacked{}$, has not been compromised. This is necessary, as full compromise would allow for full access to the API that we know to be flawed. As we explain next, we do not resort to any API modification and, instead, we base our separation of roles on the standard \wwt{} attribute (cf. \autoref{sec:rules}).

\subsection{Security Property}
\label{sec:security}

The primary aim of an HSM is to never expose sensitive keys outside the device: cryptographic operations are performed inside the HSM and keys should only be exported, when necessary, wrapped under other secure keys. Since key usage is allowed by the API, we cannot guarantee any security property about it: attacker accessing the API as user $u\in\attacked{}$ will be able to perform any operation on behalf of $u$ such as encrypting, decrypting and signing data. In fact, the latest version of \pkcs{} usage guide \cite[Section 3.1]{PKCS11usage} explicitly states that an authenticated user may perform any operation supported by the token.
%: ``Once a normal user has been authenticated to the token, \ldots the user may perform any operation supported by the token''.
So, even when the keys are secured we do not have any guarantee about data security, when the HSM is under \attacked{}-attack (cf. \autoref{def:attacker}).

Therefore, our focus is on key confidentiality: any sensitive key of the HSM should remain confidential, even when the device is under attack. One might wonder which keys should be regarded as sensitive and which should not. Typically, long term keys stored in the device and reused across sessions should be regarded as sensitive. Session keys, generated on the fly and with a limited validity might be regarded as less important and treated with more flexibility: an API-level attack compromising session keys would have a limited impact and would not allow to clone the HSM functionality. Notice that, \sensitive{} is also a \pkcs{} attribute that is used to explicitly tag sensitive keys in the device, but we use the word sensitive in its broader sense, to indicate important keys as defined below:

\begin{mydef}[Sensitive keys]
\label{def:sensitive}
Any key that, if leaked, allows to clone the HSM functionality should be regarded as \emph{sensitive} and should have the corresponding \pkcs{} attribute set.
\end{mydef}
We can now state our security property for HSMs  in terms of Definitions~\ref{def:attacker} and \ref{def:sensitive}. As discussed in \autoref{sec:attacker}, we define security assuming that 
a portion of the users, namely $\users{} \setminus \attacked{}$, has not been compromised:

\begin{mydef}[HSM security]
\label{def:security}
Given a subset of the HSM users $\attacked{} \subseteq \users{}$, we say that an HSM is \attacked{}-secure if no \emph{sensitive} key can be leaked even when the HSM is under \attacked{}-attack.
\end{mydef}

\subsection{User Roles}
\label{sec:users}

Our secure configuration is based on a careful separation of roles so that the critical key management operations are only performed by special users, in a way that production applications cannot exploit the (flawed) API to extract keys. This is very important, especially in a cloud setting, where the HSM is operated remotely and a vulnerability in one application might compromise the security of the whole device, leaking all of the sensitive keys. Interestingly, our solution does not require any limitation of the API, which makes it fully compatible with standard HSM implementations, assuming a mild form of access control over keys.
We define three categories of users:

\begin{description}

\item[Normal Users (NU)] These users are accessible by production applications, so it is of ultimate importance that: $(i)$ they can use the full API, so to maximize compatibility; $(ii)$ they cannot exploit API-level attacks to leak sensitive keys, as they are the most exposed HSM users: in fact, an attacker might exploit a vulnerability in production application to gain access to the HSM API.

\item[Key Manager (KM)] These users perform key management operations using the standard API in a controlled way. They are responsible for handling the \trusted{} keys that will be used to wrap other sensitive keys. The KM accounts should only be accessed by special management applications (or directly by humans), and their credentials should never be used in production applications. 

\item[Security Officer (SO)] The Security Officer is a special user which adheres to the \pkcs{} standard and has no access to the full API \cite[Section 2.4]{PKCS11usage}. As such, the SO mainly performs administrative tasks, such as creating other HSM users. Interestingly, the SO can mark as \trusted{} the keys generated by other users. As for KMs, the SO account should only be accessed by special management applications (or directly by humans), and its credentials should never be used by production applications. 

\end{description} 
It is important to notice that the \pkcs{} standard only defines two types of users: the normal users and the Security Officer (SO) \cite[Section 2.4]{PKCS11usage}. Thus, KMs will be implemented as a normal users whose credential should be regarded as very critical: compromising a KM would allow for API-level attacks.  In practice, the SO and the KMs could be accessed by the same physical person using different accounts and combining cryptographic and administrative operations in order to generate and manage \trusted{} keys.

\subsection{Secure Configuration}
\label{sec:rules}

We now present a secure configuration based on \trusted{} and \wwt{}  attributes, and on the user roles  defined in \autoref{sec:users}.

\subsubsection*{Protecting sensitive keys}

The only available defense against API-level attacks was introduced in version 2.40 of the \pkcs{} standard: a special attribute \wwt{} that, when set on \emph{sensitive} keys, only allows wrapping under a \trusted{} key which, in turn, can only be set by a special user of the HSM, namely the Security Officer. Not using \wwt{} completely voids the adoption of an HSM as any attacker getting access to the API can potentially leak any key which has not been marked as \wwt{} with a trivial wrap-then-decrypt attack. For this reason, we require this attribute on any sensitive key of the HSM (cf. \autoref{def:sensitive}). As an alternative, the key should be generated with the \extractable{} attribute unset, meaning that the key cannot be wrapped under other keys. Recall from \autoref{sec:pkcs}, that \wwt{} cannot be unset and \extractable{} cannot be set, so if a key is generated with the suggested values the configuration is permanent and cannot be subverted: either the key is wrapped under a  \trusted{}  key or it cannot be wrapped at all.

\begin{guideline}[Sensitive keys]
\label{g:wwt}
Any sensitive key stored in the HSM should either be generated with 
attribute \wwt{} set, or with
attribute \extractable{} unset.
\end{guideline}

\subsubsection*{Restricting KMs and SO behavior} 

In order to guarantee that \trusted{} keys can really be trusted we need to impose constraints on the process of tagging a key as \trusted{}. Recall that only the SO can set the \trusted{} attribute, while cryptographic operations, such as key generation, should be done by a normal user. We elect KM to carry out this delicate task. The idea is that only the KM can generate a key that will be promoted to \trusted{} by the SO.

\begin{guideline}[Trusted keys]
\label{g:trusted}
The SO sets attribute \trusted{} only on special \emph{candidate} keys generated by one of the KMs.
\end{guideline}

These candidate keys should be treated very carefully so to avoid possible API-level attacks in production applications. In particular, during their lifetime they should never have roles that conflict with key wrapping, such as \encrypt{} and \decrypt{}:

\begin{guideline}[Roles of candidate keys]
\label{g:roles}
The candidate keys managed by the KMs should only admit wrap and unwrap cryptographic operations during their whole lifetime.
\end{guideline}

Moreover, candidate keys should not be wrapped under other keys, to prevent attacks in which keys are exported and then reimported with different attributes. Unfortunately, this uncovers the limitation of the \wwt{} mechanism which does not allow for managing a hierarchy of keys.

\begin{guideline}[Management of candidate keys]
\label{g:no_extract}
The candidate keys managed by the KMs should be generated with the \extractable{} attribute unset.
\end{guideline}

Candidate keys should be generated in the device to prevent them to be used differently from what is required by previous rules. Recall that, imported keys might exist in other devices with conflicting attributes. Cloud HSM solutions offer mechanisms to transparently synchronize keys across cluster of devices, out of the \pkcs{} standard (e.g., \cite[page 84]{AWScloudHSM}). So, in practical implementations it might be fine to generate fresh wrapping keys that are transparently replicated in other HSMs belonging to the same cluster.

\begin{guideline}[Freshness of candidate keys]
\label{g:confidentiality}
The candidate keys managed by the KMs should be generated as fresh in the device.
\end{guideline}

\subsection{Security Analysis}
\label{sec:sec_analysis}

We now informally discuss how the configuration of \autoref{sec:rules} prevents the known API-level attacks from the literature. We limit our analysis to the attacks that exploit key management APIs to extract sensitive keys in the clear presented in \autoref{sec:pkcs}. Note that,  \autoref{g:wwt} requires to wrap sensitive keys only under \trusted{} keys which, according to  \autoref{g:trusted}, can only be KM's candidate keys. So we get the following:

\begin{fact}
\label{f:candidate}
If Rules \ref{g:wwt} and \ref{g:trusted} are respected, then sensitive keys can only be wrapped under keys that respect Rules \ref{g:roles}, \ref{g:no_extract} and \ref{g:confidentiality}.
\end{fact}
We now show how the various attacks are prevented by one of these rules on candidate keys. 

\subsubsection*{Wrap-then-decrypt} This attack is based on wrapping a sensitive key $k_1$ with a \wrap{} key $k_2$ and subsequently decrypting it. It requires that the \wrap{} key $k_2$ can be also used to decrypt data. Because of  \autoref{f:candidate}, sensitive keys can only be wrapped under keys that respect \autoref{g:roles} which, in turns, forbids decryption with $k_2$ and prevents the attack.

\subsubsection*{Encrypt-then-unwrap} This attack is based on encrypting a known attacker key $k_A$ and then importing it in the device as a malicious wrapping key. Because of  \autoref{f:candidate}, sensitive keys can only be wrapped under keys that respect \autoref{g:confidentiality} which, in turns, forbids \wrap{} keys that have not been generated in the device, such as $k_A$. This prevents the attack.

\subsubsection*{Importing with conflicting roles} A key $k_A$ is imported twice, setting \wrap{} attribute on one copy and \decrypt{} on the other so to execute a wrap-then-decrypt attack. As for the previous case, \autoref{f:candidate} and \autoref{g:confidentiality} prevent wrapping under a key $k_A$ that has not been generated in the device.

\subsubsection*{Reimporting with conflicting roles} A wrapping (\trusted{})  key $k_T$ is wrapped and then unwrapped, setting the \decrypt{} attribute on the copy, so to execute a wrap-then-decrypt attack. In this case \autoref{g:confidentiality} does not help as the key could have been generated in the device. However, \autoref{f:candidate} and \autoref{g:no_extract} prevent that the \trusted{} wrapping key $k_T$ is wrapped in the first place, blocking the attack.

\medskip
A fundamental ingredient of our configuration is that both SO and KM behave accordingly to  Rules \ref{g:trusted} and \ref{g:roles}, \ref{g:no_extract}, \ref{g:confidentiality}, respectively. Therefore, we assume that SO and KM users are  immune to attacks (cf. \autoref{sec:attacker}). NUs only have to respect \autoref{g:wwt} in order to have their sensitive keys protected by the \wwt{} mechanism, but this rule does not restrict their behavior in any way: once keys are tagged as \wwt{} these users can access the full API and those keys will be immune to API-level attacks. Recall that, we implicitly assume that only the  owner of the key can modify its attribute (cf. \autoref{def:owner}) so, for example, NUs cannot modify the attributes of KM's keys. This is fundamental to prevent trivial attacks on trusted keys.

We have shown that some of the prominent API-level attacks are blocked by the rules of our configuration and, in particular, we have seen that all of the rules are necessary in order to prevent the attacks. We claim that our configuration guarantees that the HSM is \attacked{}-secure (cf. \autoref{def:security}) when SO and KM users are not under attack, i.e., they do not belong to \attacked{}. In other words, if the attacker only accesses the HSM as a normal user, sensitive keys are never leaked even when the attacker access the full API. More precisely:

\begin{claim}[Secure configuration]
\label{claim:security}
Let \usersSO{} and  \usersKM{} respectively denote all SO and KM users. If all the configuration rules of \autoref{sec:rules} are respected then the HSM is \attacked{}-secure for each $\attacked{} \subseteq \users$ such that $\attacked \cap (\usersSO{} \cup  \usersKM{}) = \emptyset$.
\end{claim}

To substantiate our claim, in \autoref{sec:formal} we will prove that the configuration is indeed correct for an unbounded number of users, sessions and keys, on a significant core of \pkcs{}.

\subsection{Implementation on Real Cloud HSMs}
\label{sec:implementation}

We now discuss how to implement the secure configuration of \autoref{sec:rules} in real cloud HSM solutions. As far as we know, AWS CloudHSM \cite{AWSHSM} is the only one suggesting in its documentation the adoption of the \wwt{} mechanism. However, the suggestion is very general and does not refer to API-level attacks. Instead, it emphasizes the possibility offered by trusted keys of defining the so called \emph{unwrap template}, i.e., the set of attributes that the imported key will have to adhere to. However, the \emph{unwrap template} only offers a false sense of security since, once a key is unwrapped, its attributes can be changed. For this reason, we have not based our solution on this particular mechanism.

Interestingly, the AWS CloudHSM solution also provides a form of access control that is expressive enough to implement our secure configuration. 

\subsubsection*{Key sharing} The AWS CloudHSM solution allows to share keys among different HSM users through a proprietary command called \texttt{shareKey} \cite[page 140]{AWScloudHSM}. The manual explicitly states that ``Users who share the key can use the key in cryptographic operations, but they cannot ... change its attributes''. The latter part is what makes the key sharing mechanism a viable implementation of our configuration: it is enough for KMs to share the candidate keys with NUs so that they can be used as wrapping keys but can never be modified. Our attacker model, in fact,  assumes a worst-case scenario in which keys are implicitly shared with any other HSM users (cf. \autoref{sec:attacker}).

\subsubsection*{Users} The AWS CloudHSM solution defines various users, two of which, named CryptoOfficer (CO) and CryptoUser (CU), can be mapped directly to the SO and normal users of \pkcs{}. Our KM can be implemented as a CU, given that we are assuming it is a normal user whose credential should not be available to production applications.

\subsubsection*{Implementation}
We assume to have special KM users that create easy to identify candidate keys, e.g., using particular labels, that we refer to as \candidatekeys{}. The secure configuration rules of \autoref{sec:rules} can be directly instantiated in AWS CloudHSM as follows:

\begin{description}

    \item[\autoref{g:wwt}] Attribute \wwt{} should be set for any sensitive key, as already suggested in \cite[page 88]{AWScloudHSM};
    \item[\autoref{g:trusted}] The CO should set the \trusted{} attribute on \candidatekeys{} keys only;
    \item[\autoref{g:roles}] Relatively to cryptographic operations, KMs should only set \wrap{} and \unwrap{} attributes on \candidatekeys{} keys;
    \item[\autoref{g:no_extract}] Keys in \candidatekeys{} should be generated with \extractable{} attribute unset;
    \item[\autoref{g:confidentiality}] Keys in \candidatekeys{} should be generated as fresh in the HSM.  Note that, fresh keys can be shared between HSMs using external mechanisms not included in \pkcs{} \cite[page 84]{AWScloudHSM}.

\end{description}

\section{Formal analysis}
\label{sec:formal}
%!TEX root = main.tex

In this section, we formalize a significant subset of \pkcs{} in the Tamarin prover \cite{MSCB13}, and we automatically prove \autoref{claim:security} for an unbounded number of users, keys and sessions. The portion of \pkcs{} that we analyze includes symmetric key encryption and key wrapping. We model \extractable{}, \wwt{} and \trusted{} attributes, as they are crucial for the secure configuration, and we model user roles mapping them to standard \pkcs{} users, i.e., Security Officer and normal users \cite[Section 2.4]{PKCS11usage}, so that we do not need to modify in any way the standard API. 

We introduce two simplifications in our model that make the analysis more manageable and fully automated. First, instead of explicitly modeling the encryption and decryption functionalities we leak keys with attribute \encrypt{} or \decrypt{} set to the attacker. Notice that, this does not reduce the attacker capabilities: on the contrary, it gives more power to the attacker since unsetting \encrypt{} and \decrypt{} will not disable the use of those keys. In our experiments, we found that this simplification did not introduce false attacks but allowed us to automatically prove all the lemmas. 
The second simplification is related to the fact that we do not model wrap and unwrap operations with untrusted keys, as we know they are subject to existing API-level attacks. In order to implicitly cover those attacks we just leak any \extractable{} key that does not have attribute \wwt{} set. Intuitively, this is equivalent to attacking each of those keys with, e.g., a wrap-then-decrypt attack.

The section is organized as follows: we describe the model of user roles, keys and attributes in \autoref{sec:model:keys}; we  discuss model simplifications in \autoref{sec:model:simplification}; we present the model of key management operations in \autoref{sec:model:wrap} and in \autoref{sec:model:analysis} we model our HSM configuration and prove its security.

\begin{figure}
\begin{lstlisting}
rule NewNU:
  [ Fr(U) ] --[ NewNU(U) ]-> [ !NU(U) ]

rule NewKM: 
  [ Fr(U) ] --[ NewKM(U) ]-> [ !NU(U) ]

rule NewSO:
  [ Fr(U) ] --[ NewSO(U) ]-> [ !SO(U) ]
\end{lstlisting}
\caption{Rules for new users.}
\label{rules:users}
\end{figure}

\subsection{Modeling User Roles, Keys and Attributes}
\label{sec:model:keys}

\subsubsection*{User roles} Standard \pkcs{} normal users and Security Officer are modeled with facts \facts{!NU(U1)} and \facts{!SO(U2)}, respectively. As explained in \autoref{sec:tamarin}, the \facts{!} symbol makes a fact persistent. We will use these facts in the precondition of rules that can be fired by such users. \autoref{rules:users} reports, directly in the Tamarin syntax, the three rules for generating the three categories of users NU, KM and SO (cf. \autoref{sec:users}). Recall that \facts{Fr} is a special fact in Tamarin used to generate fresh names that are not known by the attacker (cf. \autoref{sec:tamarin}). For example, rule \facts{NewNU} generates the persistent fact \facts{!NU(U)}, for a fresh username \facts{U}, producing an action fact \facts{NewNU(U)}. This action fact appears in the protocol execution trace and can be used in logical formulas to state theorems over the model execution. Freshness of name \facts{U} ensures that each user has a unique role. Note that, rule \facts{NewKM} also generates a normal user \facts{!NU(U)} but the action is different: \facts{NewKM(U)}. This allows KMs to use the full API as if they were NUs, but the different action will allow us to set special restrictions over KMs (cf. \autoref{sec:model:analysis}).

\begin{figure}
\begin{lstlisting}
rule CreateKey:
  [ !NU(U), Fr(ha), Fr(k) ] 
--[ CreateKey(U,ha,k), 
    SetAttr(U,ha,'extractable') ]->
  [ !Key(U,ha,k) ]

rule CreateWWTKey:
  [ !NU(U), Fr(ha), Fr(k) ] 
--[ CreateWWTKey(U,ha,k), 
    SetAttr(U,ha,'wrap_with_trusted'), 
    SetAttr(U,ha,'extractable') ]->
  [ !Key(U,ha,k) ]

rule CreateNEKey:
  [ !NU(U), Fr(ha), Fr(k) ] 
--[ CreateNEKey(U,ha,k) ]->
  [ !Key(U,ha,k) ]

rule ImportKey:
  [ !NU(U), Fr(ha), In(k) ] 
--[ ImportKey(U,ha,k), 
    SetAttr(U, ha, 'extractable') ]->
  [ !Key(U,ha,k) ]
\end{lstlisting}
\caption{Rules for key creation and import.}
\label{rules:keys}
\end{figure}

\subsubsection*{Cryptographic keys} Recall that \pkcs{} uses handles to refer to cryptographic keys that are stored in the device (cf. \autoref{sec:pkcs}). According to \autoref{def:owner}, we also need to store the owner of the key. Therefore, we model keys as a persistent fact \facts{!Key(U,ha,k)} with three terms: the owner \facts{U}, the handle \facts{ha} and the key value \facts{k}. In \pkcs{} it is possible to specify a \emph{template} that assigns values to attributes when a key is created/imported. The most generic template is one that only sets attribute \extractable{}. Recall, in fact, that this attribute cannot be set but only unset, so if a key is generated with \extractable{} unset it will never be exportable. In principle, from this template it is possible to reach any possible combination of attributes since, as we will describe next, attributes can be set and unset (with some exceptions). However, keys generated with this generic template are subject to known API-level attacks, so we need to define two more templates that we will use to generate secure keys. In fact, these templates adhere to \autoref{g:wwt}: The former has both \wwt{} and \extractable{} set and the latter has attribute \extractable{} unset.

Key creation/import rules are reported in \autoref{rules:keys}. The first three rules correspond to the three templates previously described and generate a new key \facts{!Key(U,ha,k)} respectively producing actions \facts{CreateKey(U,ha,k)}, \facts{CreateWWTKey(U,ha,k)} and \facts{CreateNEKey (U,ha,k)}. Setting an attribute is modeled by \facts{SetAttr(U,ha,a)}, representing user \facts{U} setting attribute \facts{a} on handle \facts{ha}. In particular, \facts{a} is a label containing the attribute name. For example, the first rule creates a key with the generic template in which only the attribute \extractable{} is set, written \facts{SetAttr(U,ha,'extractable')}; similarly, the second rule creates a key with  \wwt{} set (\extractable{} is also set in order to make the key exportable under wrapping); note that,  the third rule does not set any attribute: not setting \extractable{} makes the key not exportable even through wrap operations. All three rules have the same precondition facts \facts{!NU(U)}, \facts{Fr(ha)}, \facts{Fr(k)}, requiring a normal user \facts{U} and two fresh values \facts{ha} and \facts{k}. The fourth  rule is for importing keys in the clear. Notice that, \facts{k} in this case is read as input through the special fact \facts{In(k)}. These keys are imported with the most generic, insecure template. In fact, this rule lets the attacker import any known key in the device and use it in subsequent attacks.

\begin{figure}
\begin{lstlisting}
// For A in {'wrap','unwrap','encrypt','decrypt'}
rule SetAttrA:
  [ !Key(U,ha,k), !NU(U) ] --[ SetAttr  (U,ha,A) ]-> [ ]
rule UnsetAttrA:
  [ !Key(U,ha,k), !NU(U) ] --[ UnsetAttr(U,ha,A) ]-> [ ]

rule UnsetAttrExtractable:
  [ !Key(U,ha,k), !NU(U) ] 
--[ UnsetAttr(U,ha,'extractable') ]-> [ ]

rule SetAttrWrapWithTrusted:
  [ !Key(U,ha,k), !NU(U) ] 
--[ SetAttr(U,ha,'wrap_with_trusted') ]-> [ ]

rule SetAttrTrusted:
  [ !Key(U,ha,k), !SO(W) ]
--[ SetAttr  (W,ha,'trusted') ]-> [ ]
rule UnsetAttrTrusted:
  [ !Key(U,ha,k), !SO(W) ] 
--[ UnsetAttr(W,ha,'trusted') ]-> [ ]
\end{lstlisting}
\caption{Rules for changing attributes. The first two rules apply to all attributes \wrap{}, \unwrap{}, \encrypt{} and \decrypt{}.}
\label{rules:attributes}
\end{figure}
\subsubsection*{Attributes} Similarly to attribute setting, attribute unsetting is modeled with action \facts{UnsetAttr(U,ha,a)}. For each attribute we add a rule for setting and one for unsetting except for \extractable{} and \wwt{} that, respectively, can only be unset and set. In \autoref{rules:attributes} we report the rules for attribute change. Apart from \trusted{}, the precondition is always \facts{Key(U,ha,k)}, \facts{!NU(U)}, i.e., the key is owned by the normal user \facts{U} who is setting/unsetting the attribute. For attribute \trusted{}, instead, the user performing the operation is SO who is not the key owner. In fact, only the Security Officer can set this attribute but cannot create the keys, that will necessarily be owned by other users.

Notice that, attributes are bound to handles and not directly to keys. This is very important and reflects the actual \pkcs{} specification: the same key can appear in the device many times under different handles, for example after many unwrap operations, and each occurrence of the key can have different attributes. This is what enables attacks in which a key is reimported with conflicting roles, which are very hard to spot in practice (cf. \autoref{sec:pkcs}). 

Action \facts{IsSet(h,a)} is a special action that we use to check that attribute \facts{a} is set on key with handle \facts{h}. In Tamarin this can be done by imposing a restriction over the action as follows:

\begin{lstlisting}
restriction IsSet:
  "
  All ha a #i . IsSet(ha,a)@i ==> 
  (
    Ex U #j . SetAttr(U,ha,a)@j & j<i &
    (All W #w . UnsetAttr(W,ha,a)@w & w<i ==> w<j)
  )
  "
\end{lstlisting}
In the restriction we universally quantify over all handles \facts{ha}, attributes \facts{a} and timepoints \facts{i} (\facts{\#} is used to indicate that \facts{i} is a time variable). Intuitively, an attribute \facts{a} is set to handle \facts{ha} at time \facts{i}, written \facts{IsSet(ha,a)@i} if the attribute has been set before, written \facts{SetAttr(U,ha,a)@j \& j<i}, and in case the attribute has been also unset before, written \facts{UnsetAttr(W,ha,a)@w \& w<i}, then the set operation happened after the unset operation, written \facts{w<j}. Intuitively, if there are many occurrences of set and unset operations before time \facts{i}, it must be that set is the last one. In fact, an attribute can be set and unset many times but what matters is the last assignment. Similarly, we define an action \facts{IsUnset} that will be used to check when an attribute is not set. The only difference with respect to \facts{IsSet} is that an attribute is unset also when it has never be set, i.e., we assume that attributes are unset by default.\footnote{Another possibility for modeling attributes would be to use non-permanent facts, but this would require re-adding an attribute each time it is used with the effect of over-complicating the model and the automated analysis.} This is formalized in the last part of the formula stating that no \facts{SetAttr(U,ha,a)@j} with \facts{j<i} exists:

\begin{lstlisting}
restriction IsUnset:
  "
  All ha a #i . IsUnset(ha,a)@i ==> 
  (
    (Ex U #j . UnsetAttr(U,ha,a)@j & j<i &
      (All W #w . SetAttr(W,ha,a)@w & w<i ==> w<j))
    |
    (not Ex U #j . SetAttr(U,ha,a)@j & j<i)
  )
  "
\end{lstlisting}
\begin{figure}
\begin{lstlisting}
rule LeakEncKey:
  [ !Key(U,ha,k) ]
--[ IsSet(ha,'encrypt') ]->
  [ Out(h(k)) ]

rule LeakDecKey:
  [ !Key(U,ha,k) ]
--[ IsSet(ha,'decrypt') ]->
  [ Out(h(k)) ]

rule LeakExtractable:
  [ !Key(U,ha,k) ]
--[ IsSet(ha,'extractable'), 
    IsUnset(ha,'wrap_with_trusted') ]->
  [ Out(h(k)) ]
\end{lstlisting}
\caption{Model simplification through explicit key leakage.}
\label{rules:leak}
\end{figure}

\subsection{Model Simplification}
\label{sec:model:simplification}

As previously discussed, instead of explicitly modeling encryption and decryption APIs we leak keys with attributes \encrypt{} or \decrypt{} set to the attacker. Moreover, since we are not interested in rediscovering known attacks, we only model wrapping under trusted keys and incorporate everything else in the standard Dolev-Yao attacker model offered by Tamarin. In fact, we do not need to model untrusted wrapping in detail because we know that it is insecure by construction. More precisely, we make the following simplifications:

\begin{itemize}

\item Keys with attributes \encrypt{} or \decrypt{} set are directly available to the attacker; as we already discussed, the attacker accessing the API can freely use the HSM keys and perform any encryption and decryption of her choice. We found out that modeling encrypt and decrypt operations complicated a lot the automated reasoning of the tool as the same term could be obtained both from the implicit Dolev-Yao attacker and from the explicit API.

\item Keys with  \extractable{} set and \wwt{} unset are directly available to the attacker. These keys are subject to a wrap-then-decrypt attack. Making them directly available to the attacker makes it possible to only model trusted wrap and unwrap functionalities.
\end{itemize}
Note that, leaking some of the keys to the attacker makes that attacker model strictly more powerful: any proof in this model will also hold in a model where those keys are not leaked. However,  in order to express and prove the secrecy lemmas (cf. \autoref{sec:model:analysis}) we need a way to distinguish between keys implicitly leaked by the model and keys leaked in actual attacks. In other words, we need to distinguish the actual key from its \emph{value}, used to perform cryptographic operations. Of course, knowing a key should allow to know its value but not vice-versa. A natural way to implement this idea is to hash  the key \facts{k} under a one-way hash function, written \facts{h(k)}, any time it is used as cryptographic key but not when it is used as data, e.g., when it is wrapped. For example in \autoref{sec:model:wrap} (\autoref{rules:wrap}) we will use the symmetric encryption term \facts{senc(k1,h(k2))} to represent key \facts{k1} wrapped under \facts{k2}, in which only the actual cryptographic key \facts{k2} is hashed. Notice that, leaking \facts{h(k)} allows the attacker to perform any cryptographic operation based on key \facts{k} but does not reveal \facts{k}. This modeling of \pkcs{} is new and it is an original contribution by itself.

In \autoref{rules:leak} we show the rules that leak keys: rules \facts{LeakEncKey} and \facts{LeakDecKey} leak respectively encryption and decryption keys, i.e., keys with \encrypt{} and \decrypt{} set. Rule \facts{LeakExtractable} leaks extractable keys that do not have \wwt{} set. 

\begin{figure}
\begin{lstlisting}
rule Wrap:
  [ !NU(U),!Key(U1,ha1,k1),!Key(U2,ha2,k2) ] 
--[ Wrap(U,ha1,ha2),
    IsSet(ha1,'wrap_with_trusted'), 
    IsSet(ha1,'extractable'), 
    IsSet(ha2,'trusted'),
    IsSet(ha2,'wrap') ]-> 
  [ Out(senc(k1,h(k2))) ]

rule Unwrap:
  [ !NU(U1),!Key(U2,ha2,k2),In(senc(k1,h(k2))),Fr(ha1) ] 
--[ Unwrap(U1,ha1,ha2),
    IsSet(ha2,'trusted'), 
    IsSet(ha2,'unwrap'),
    SetAttr(U1,ha1,'wrap_with_trusted'), 
    SetAttr(U1,ha1,'extractable') ]-> 
  [ !Key(U1,ha1,k1) ] 
\end{lstlisting}
\caption{Wrap and unwrap of \wwt{} keys.}
\label{rules:wrap}
\end{figure}

\subsection{Modeling Key Management}
\label{sec:model:wrap}

As already mentioned, we are not interested in rediscovering known API-level attacks so we explicitly model 
% we limit our attention to 
wrap and unwrap operations with trusted keys, while we implicitly leak keys with attribute \wwt{} unset (cf. \autoref{sec:model:simplification}).  In \autoref{rules:wrap}, we model wrap and unwrap operations in which the wrapped key has attributes \wwt{} and \extractable{} set and the wrapping key has \trusted{} and \wrap{}/\unwrap{} set. Intuitively, in rule \facts{Wrap} a normal user \facts{U} wraps key \facts{k1} belonging to user \facts{U1} under key \facts{k2} belonging to user \facts{U2}. Handle \facts{ha1} of \facts{k1} has both \wwt{} and \extractable{} set while handle \facts{ha2} of \facts{k2} has both \trusted{} and \wrap{} set. The encryption of \facts{k1} under \facts{k2}, noted \facts{senc(k1,h(k2))}, is sent as output using the special Tamarin fact \facts{Out}.
Similarly, in rule \facts{Unwrap} a normal user \facts{U1} unwraps key \facts{k1} from the ciphertext \facts{senc(k1,h(k2))} read as input through the special Tamarin fact \facts{In}. The key is imported under a fresh handle \facts{ha1}. The wrapping key is required to have both \trusted{} and \unwrap{} set and the new key is imported with both \wwt{} and \extractable{} set. The reason  why we use \facts{h(k2)} instead of \facts{k2} is due to the model simplification explained in \autoref{sec:model:simplification}.

\subsection{Proof of Security}
\label{sec:model:analysis}
We now restrict the behavior of SO and KMs to adhere to the secure configuration of \autoref{sec:rules}. 

\subsubsection*{Trusted keys} \autoref{g:trusted} states that SO should only mark as \trusted{} special candidate keys generated by KMs. We let these keys be all the ones generated with \extractable{} unset (rule \facts{CreateNEKey} of \autoref{rules:keys}). This also complies with \autoref{g:confidentiality}, since the key is generated as fresh, and with \autoref{g:no_extract}, since the key is generated with \extractable{} unset. We prove security for all of these keys but, in a real implementation, candidate keys can be a subset of them.

\begin{lstlisting}
restriction SO:
  "
  All W ha #i . SetAttr(W,ha,'trusted')@i
  ==>
  Ex k U #j #w . 
    CreateNEKey(U,ha,k)@j & j<i &
    NewKM(U)@w & w<i
  "
\end{lstlisting}
The restriction states that whenever a handle \facts{ha} is marked as trusted by \facts{W} at time \facts{i}, written \facts{SetAttr(W,ha,'trusted')@i}, then the key must have been created with \extractable{} unset, written \facts{CreateNEKey(U,ha,k)@j \& j<i}, by a Key Manager \facts{U}, written \facts{NewKM(U)@w \& w<i}.

\subsubsection*{Candidate keys} Rules \ref{g:roles}, \ref{g:no_extract} and \ref{g:confidentiality} state how KMs should handle candidate keys that are marked as \trusted{} by SO. \autoref{g:confidentiality} and \ref{g:no_extract} are already fulfilled by the fact that SO only marks as trusted freshly generated keys with \extractable{} unset. As for Rule \ref{g:roles}, we only need to require that KM only admits wrap and unwrap cryptographic operations on these keys, obtaining the following restriction on KM's behavior:

\begin{lstlisting}
restriction KM:
  "
  All U ha k a #i #j #w.  
    NewKM(U)@i &
    CreateNEKey(U,ha,k)@j &
    SetAttr(U,ha,a)@w
  ==> 
    ( a='wrap' | a='unwrap' )
  "
\end{lstlisting}
The restriction states that whenever an attribute \facts{a} is set by a Key Manager \facts{U} on a non \extractable{} key \facts{k} with handle \facts{ha}, the attribute can only be \wrap{} or \unwrap{}. Note that, there is no constraint on the timing of actions, so this restriction holds independently of the ordering of actions for the entire lifetime of the key, as required by \autoref{g:roles}.

\subsubsection*{HSM security}
We now instantiate and prove \autoref{claim:security} on the Tamarin model presented so far that we have intuitively shown to adhere to the secure configuration rules of \autoref{sec:configuration}. \autoref{claim:security} states that if all the rules are respected then the HSM is \attacked{}-secure when any user except SO and KMs are under the control of the attacker. This assumption is implemented in the model by the restrictions over SO and KM  presented above. From \autoref{def:security} we know that \attacked{}-secure means that no \emph{sensitive} key is leaked. Trusted keys should be certainly regarded as sensitive, as they are used to wrap other sensitive keys. Moreover, by \autoref{g:wwt} we require that sensitive keys should either  be generated with attribute \wwt{} set, or with attribute \extractable{} unset, i.e., with \facts{CreateWWTKey} or \facts{CreateNEKey} of \autoref{rules:keys}. In the following, we prove that the confidentiality of all of these keys is preserved.

For expressing the secrecy of trusted key it is useful to define an action \facts{IsHandle(ha,k)} stating that \facts{ha} is a valid handle for key \facts{k}. We add a specific rule producing this action for each key:
\begin{lstlisting}
rule IsHandle:
  [ !Key(U,ha,k) ] --[ IsHandle(ha,k) ]-> [ ]
\end{lstlisting} 
Now, we can state the secrecy of trusted keys as follows:
\begin{lstlisting}
lemma SecrecyTrusted:
  "
  All W ha k #i #j #w. 
    IsHandle(ha,k)@i & 
    SetAttr(W,ha,'trusted')@j & 
    KU(k)@w 
  ==> F
  "
\end{lstlisting}
where \facts{KU(k)} is a special Tamarin fact expressing that the attacker knows \facts{k} and \facts{F} denotes \emph{false}. This lemma states that any key \facts{k} with handle \facts{ha} marked as \trusted{} cannot be discovered by the attacker (implication of \emph{false} requires that the hypothesis is \emph{false}).

The secrecy of \facts{CreateWWTKey} and \facts{CreateNEKey} can be stated in a similar way:

\begin{lstlisting}
lemma SecrecyNE:
  "
  All U ha k #i #j . 
    CreateNEKey(U,ha,k)@i & 
    KU(k)@j 
  ==> F
  "

lemma SecrecyWWT:
  "
  All U ha k #i #j . 
    CreateWWTKey(U,ha,k)@i & 
    KU(k)@j 
  ==> F
  "
\end{lstlisting}
These three lemmas constitute an instance of \autoref{claim:security} and can be automatically checked  in Tamarin, as detailed in the next section.

\subsection{Automated Verification in Tamarin}
The complete Tamarin model is in \autoref{appendix} and is publicly available at \cite{CloudHSM-model}. In addition to what was presented in the previous sections, it includes a helper lemma used to make the proof converge and a number of sanity lemmas checking that: $(i)$ the rules of \autoref{sec:rules} are correctly modeled; $(ii)$ all the actions are executable, i.e., the model is actually executable; $(iii)$ some intuitive facts hold. 

\subsubsection*{Helper lemma} Lemma \facts{Unwrap} is a so called \emph{source} lemma. Tamarin cannot compute all the possible sources of terms used in its backward search algorithm for our model and this lemma helps Tamarin to disambiguate some cases. In particular the lemma states that unwrapped keys come from one of the \facts{CreateKey}, \facts{CreateWWTKey} or \facts{ImportKey} rules:

\begin{lstlisting}
lemma Unwrap [sources, reuse]:
  "
  All U k1 ha1 ha2 #i . 
    Unwrap(U,ha1,ha2)@i & 
    IsHandle(ha1,k1)@i
  ==> 
  (
    (Ex W ha #j . CreateKey(W,ha,k1)@j & j<i)
    |
    (Ex W ha #j . CreateWWTKey(W,ha,k1)@j & j<i)
    |
    (Ex W ha #j . ImportKey(W,ha,k1)@j & j<i)
  )
  "
\end{lstlisting}
Notice that, \facts{CreateNEKey} is not possible as it generates non extractable keys. Once this lemma is proved all the sources are correctly computed and other lemmas can be proved based on this one because of the \facts{reuse} label.

\subsubsection*{Sanity lemmas for rules} We check the following properties for the rules of  \autoref{sec:rules}:

\begin{description}
  \item[\autoref{g:wwt}] Sensitive keys, i.e., keys that we proved to remain confidential in \autoref{sec:model:analysis}, have been either generated with attribute \wwt{} set (\facts{SanityRule1\_1}), or with attribute \extractable{} unset (\facts{SanityRule1\_2} and \facts{1\_3}). In the lemmas, we additionally check that these attributes cannot respectively be unset or set;

  \item[\autoref{g:trusted}] Only the SO can set the \trusted{} attribute and the candidate keys are always generated by a KM with \extractable{} unset (\facts{SanityRule2\_1} and \facts{2\_2});

  \item[\autoref{g:roles}] The candidate keys managed by the KMs should only admit \wrap{} and \unwrap{} cryptographic operations during their whole lifetime (\facts{SanityRule3});

  \item[\autoref{g:no_extract}] The candidate keys have the \extractable{} attribute unset 
 during their whole lifetime (\facts{SanityRule4}), which implies that they have been generated with the \extractable{} attribute unset.
\end{description}
For what concerns {\autoref{g:confidentiality}}, the fact that the candidate keys managed by the 
 KMs are generated as fresh in the device trivially derives from rule \facts{CreateNEKey} that requires
 \facts{Fr(k)} in the premise. So for this rule we do not prove any sanity lemma.

\subsubsection*{Sanity lemmas for actions} It is important to check that the model really 
 does something and check that all the actions are possible. These sanity lemmas are all of the form \facts{exists-trace} as they look for one trace satisfying the required formula. For example:

 \begin{lstlisting}
lemma SanityUsers:
exists-trace
"
Ex U1 U2 U3 #i1 #i2 #i3.
  NewSO(U1)@i1 & NewNU(U2)@i2 & NewKM(U3)@i3 
"
\end{lstlisting}
checks that at least one user for each role can be created. We have similar lemmas for all the actions defined in the model.

\subsubsection*{Sanity lemmas for intuitive facts} We finally check some intuitive facts that are expected to hold: 
\begin{description}
 \item[\rm\facts{SanityUsersRole}] checks that the same user can never have two distinct roles;  
 \item[\rm\facts{SanityAttributesExtractable2}] checks that the attribute \linebreak \extractable{}  can be set only by  \facts{CreateKey}, \facts{CreateWWTKey}, \linebreak \facts{ImportKey} and \facts{Unwrap}; 
 \item[\rm\facts{SanityAttributesWWT2}] checks that \wwt{} attribute can never be unset.
\end{description}
All lemmas can be proved automatically in about 1m30s on a MacBook Pro 2018. The full model and more detail about its automated verification are available at \cite{CloudHSM-model}.

\section{Conclusion}
\label{sec:conclusion}
%!TEX root = main.tex

\pkcs{} is notoriously vulnerable and existing fixes require non-standard mechanisms and/or a reduced functionality. HSMs are increasingly offered in the cloud and they must adhere to \pkcs{} in order to offer a uniform and standard interface to applications. This makes proprietary fixes even less appealing and practical. Motivated by this urgency to configure HSMs in a secure way without loosing compliance with the \pkcs{} standard, we have proposed the first practical HSM configuration that does not require any restrictions or modifications of the \pkcs{} API. 

In particular, we have explored for the first time users' roles for securing \pkcs{} by assuming a mild form of access control that only allows key owners to change key attributes. Based on this idea, we have defined five rules that we claim to be sufficient to prevent API-level attacks in \pkcs{} devices, preserving key confidentiality. We have substantiated our claim by providing a formal model of a significant subset of \pkcs{} and we have proved the security of the proposed rules on an unbounded number of users, keys and sessions. The model is expressive enough to cover all known API-level attacks, so the proof of security is promising and supports the claim that the proposed rules enforce security on the entire API. The proof is mechanized in Tamarin making it possible to easily verify further variants of the proposed configuration.

Interestingly, the simple access control underneath our configuration is offered by AWS CloudHSM and we have shown how our rules can be directly mapped and implemented in this real cloud HSM solution. In fact, the proposed configuration is rather simple to implement and one might wonder if more sophisticated configurations exist. 
For example, we might try to relax \autoref{g:confidentiality} so that existing keys can be imported and used in the device as \trusted{} keys. Unfortunately, this cannot be done in \pkcs{} as the only way to import a key securely is through unwrap and we know that unwrap can be abused to generate instances of the same key with conflicting attributes. 

The simplicity of our solution in a sense reflects the simplicity of the wrap-with-trusted mechanism which unfortunately is not suitable 
for managing a key hierarchy in which the wrapping keys can themselves be wrapped by keys that are higher in the hierarchy (see, e.g., \cite{CORTIERGeneric}).
Note, in fact, that \autoref{g:wwt} states that any sensitive key should either be generated with 
attribute \wwt{} set, or with
attribute \extractable{} unset. Trusted keys are certainly sensitive and cannot be wrapped and unwrapped to prevent changes of attributes so the only possibility is to generate them with \extractable{} unset (cf. \autoref{g:no_extract}), meaning that they cannot be wrapped under other keys.

As a future work we intend to investigate if more sophisticated access control policies on keys can help building more complex and flexible secure configurations of \pkcs{}. Moreover, we want to explore how to implement our configuration in other cloud HSM solutions.
From a preliminary study we have seen that, unfortunately, some solutions such as Utimaco \cite{UtimacoHSM} and Microsoft \cite{MicrosoftHSM} do not have a publicly available technical documentation. IBM Cloud HSM \cite{IBMHSM} implements an access management mechanism called Cloud Identity and Access Management (IAM) that regulates access to cryptographic operations. However, the form of key sharing required by our solution is only mentioned for key rings which do not support \pkcs{} keys \cite{IBMHSMKeyRings}. Understanding whether our solution could be implemented in IBM Cloud HSM requires more investigation. Surprisingly, Google Cloud HSM documentation \cite{GoogleHSM} does not mention \pkcs{} at all. From discussions in forums it seems like it is not (yet) supported.
Finally, we plan to contact AWS and Oasis to discuss possible collaborations in terms of real implementations and standard inclusion.

\begin{acks}
We would like to thank the anonymous reviewers for their very interesting and insightful comments. This work has been partially supported by the European Regional Development Fund project \emph{SAFE PLACE: Sistemi IoT per ambienti di vita salubri e sicuri} (POR FESR 2014-2020  AZIONE 1.1.4 DGR 822/2020 --- ID 10288513).
\end{acks}

\bibliographystyle{ACM-Reference-Format}
\bibliography{ms}

\appendix
\section{The Tamarin Model}
\label{appendix}
%!TEX root = main.tex

% \subsubsection*{The Tamarin model}
The complete model in Tamarin is listed below and is publicly available at \cite{CloudHSM-model}:
\begin{lstlisting}[basicstyle=\tiny\ttfamily,frame=single]
/*
 * Tamarin model for the ACM CCS'21 paper:
 * 
 * A Formally Verified Configuration for Hardware Security Modules in 
 * the Cloud by R. Focardi and F. L. Luccio.
 * 
 * Rationale: we model the trusted part and keep everything else Dolev-Yao. 
 * We do not model untrusted keys in detail because we know they are 
 * insecure-by-construction. Modeling just the trusted part of the API 
 * simplifies the model and is a new contribution by itself.
 *
 * Some keys are leaked explicitly:
 *
 * - encrypt/decrypt keys (they can be used arbitrarily);
 *
 * - extractable and non wrap_with_trusted keys, as they are subject to 
 *   known API-level attacks.
 *
 * Attacks found here are attacks in the full model. In order to distinguish
 * between a key and its value we actually leak a one-way hash of the key
 * and we always use the hash to perform cryptographic operations.
 * 
 * Check with: 
 * tamarin-prover --prove HSM_model_CCS_cameraready.spthy
 * (about 1m30s on a MacBook pro 2018)
 *
 * Check only the secrecy lemmas with:
 * tamarin-prover --prove=Secrecy* HSM_model_CCS_cameraready.spthy
 * (about 22s on a MacBook pro 2018)
 */

theory HSM_model_CCS

begin

builtins: symmetric-encryption, hashing

/*
 * User roles:
 * 
 * NU: Normal user (for production applications)
 * KM: Key Manager (implemented as a normal PKCS11 user)
 * SO: Security Officer
 *
 * NOTE: users are fresh names to prevent role clashes. See also
 * lemma SanityUsersRole below.
 */
rule NewNU:
  [ Fr(U) ] --[ NewNU(U) ]-> [ !NU(U) ] // Normal User

rule NewKM: 
  [ Fr(U) ] --[ NewKM(U) ]-> [ !NU(U) ] // Key Manager, a special NU

rule NewSO:
  [ Fr(U) ] --[ NewSO(U) ]-> [ !SO(U) ] // Security Officer

/*
 * Key creation and import.
 *
 * NOTE: Keys are created/imported with extractable set, 
 * since it cannot be set later on, only unset.
 *
 * We also need keys generated: 
 * - with extractable unset, that will become trusted (CreateNEKey)
 * - with wrap_with_trusted set, that will remain secure (CreateWWTKey)
 */
rule CreateKey:
  [ !NU(U), Fr(ha), Fr(k) ] 
--[ CreateKey(U,ha,k), 
    SetAttr(U,ha,'extractable') ]->
  [ !Key(U,ha,k) ]

rule CreateWWTKey:
  [ !NU(U), Fr(ha), Fr(k) ] 
--[ CreateWWTKey(U,ha,k), 
    SetAttr(U,ha,'wrap_with_trusted'), 
    SetAttr(U,ha,'extractable') ]->
  [ !Key(U,ha,k) ]

rule CreateNEKey:
  [ !NU(U), Fr(ha), Fr(k) ] 
--[ CreateNEKey(U,ha,k) ]->
  [ !Key(U,ha,k) ]

rule ImportKey:
  [ !NU(U), Fr(ha), In(k) ] 
--[ ImportKey(U,ha,k), 
    SetAttr(U, ha, 'extractable') ]->
  [ !Key(U,ha,k) ]

/*
 * This is never used in the model but just in lemmas/restrictions.
 * In some lemmas we need to check the mapping handle->key
 */
rule IsHandle:
  [ !Key(U,ha,k) ] --[ IsHandle(ha,k) ]-> [ ]

/*
 * Attributes
 *
 * - extractable: key is extractable, can only be unset;
 * - wrap: key can wrap other keys;
 * - unwrap: key can unwrap other keys;
 * - encrypt: key can encrypt data;
 * - decrypt: key can decrypt data;
 * - wrap_with_trusted: key can be wrapped under trusted keys,
 *   can be set but not unset.
 *   Note that we only model trusted wrapping so these are the
 *   only keys that are wrapped in the model;
 * - trusted: keys that can wrap wrap_with_trusted ones.
 *   These can only be set by the SO.
 */
rule UnsetAttrExtractable:
  [ !Key(U,ha,k), !NU(U) ] --[ UnsetAttr(U,ha,'extractable') ]-> [ ]

rule SetAttrWrap:
  [ !Key(U,ha,k), !NU(U) ] --[ SetAttr  (U,ha,'wrap')   ]-> [ ]
rule UnsetAttrWrap:
  [ !Key(U,ha,k), !NU(U) ] --[ UnsetAttr(U,ha,'wrap')   ]-> [ ]

rule SetAttrUnwrap:
  [ !Key(U,ha,k), !NU(U) ] --[ SetAttr  (U,ha,'unwrap') ]-> [ ]
rule UnsetAttrUnwrap:
  [ !Key(U,ha,k), !NU(U) ] --[ UnsetAttr(U,ha,'unwrap') ]-> [ ]

rule SetAttrEncrypt:
  [ !Key(U,ha,k), !NU(U) ] --[ SetAttr  (U,ha,'encrypt') ]-> [ ]
rule UnsetAttrEncrypt:
  [ !Key(U,ha,k), !NU(U) ] --[ UnsetAttr(U,ha,'encrypt') ]-> [ ]

rule SetAttrDecrypt:
  [ !Key(U,ha,k), !NU(U) ] --[ SetAttr  (U,ha,'decrypt') ]-> [ ]
rule UnsetAttrDecrypt:
  [ !Key(U,ha,k), !NU(U) ] --[ UnsetAttr(U,ha,'decrypt') ]-> [ ]

rule SetAttrWrapWithTrusted:
  [ !Key(U,ha,k), !NU(U) ] --[ SetAttr(U,ha,'wrap_with_trusted') ]-> [ ]

rule SetAttrTrusted:
  [ !Key(U,ha,k), !SO(W) ] --[ SetAttr  (W,ha,'trusted') ]-> [ ]
rule UnsetAttrTrusted:
  [ !Key(U,ha,k), !SO(W) ] --[ UnsetAttr(W,ha,'trusted') ]-> [ ]

/* 
 * Trusted wrap: Key(U1,ha1,k1) can be wrapped under Key(U2,ha2,k2) if
 * - ha1 has wrap_with_trusted and extractable set;
 * - ha2 has trusted and wrap set.
 * The ciphertext senc(k1,k2) is sent as output.
 */
rule Wrap:
  [ !NU(U), !Key(U1,ha1,k1), !Key(U2,ha2,k2) ] 
--[ Wrap(U,ha1,ha2),
    IsSet(ha1,'wrap_with_trusted'), 
    IsSet(ha1,'extractable'), 
    IsSet(ha2,'trusted'),
    IsSet(ha2,'wrap') ]-> 
  [ Out(senc(k1,h(k2))) ]

/*
 * Trusted unwrap: Key(U1,ha1,k1) can be unwrapped under Key(U2,ha2,k2) if
 * ha2 has trusted and unwrap set.
 * The new key Key(U1,ha1,k1) has wrap_with_trusted and extractable set.
 */
rule Unwrap:
  [ !NU(U1), !Key(U2,ha2,k2), In(senc(k1,h(k2))), Fr(ha1) ] 
--[ Unwrap(U1,ha1,ha2),
    IsHandle(ha1,k1), // used in lemma
    IsSet(ha2,'trusted'), 
    IsSet(ha2,'unwrap'),
    SetAttr(U1, ha1, 'wrap_with_trusted'), 
    SetAttr(U1, ha1, 'extractable') ]-> 
  [ !Key(U1,ha1,k1) ] 

/*
 * We leak (a hash of) encrypt/decrypt key so that attacker does 
 * any crypto operation.
 */
rule LeakEncKey:
  [ !Key(U,ha,k) ]
--[ IsSet(ha,'encrypt') ]->
  [ Out(h(k)) ]

rule LeakDecKey:
  [ !Key(U,ha,k) ]
--[ IsSet(ha,'decrypt') ]->
  [ Out(h(k)) ]

/*
 * Non wrap_with_trusted keys that are extractable are implicitly 
 * compromised.
 */
rule LeakExtractable:
  [ !Key(U,ha,k) ]
--[ IsSet(ha,'extractable'), 
    IsUnset(ha,'wrap_with_trusted') ]->
  [ Out(h(k)) ]

/*
 * An attribute a is set over ha, written IsSet(ha,a) if there exist 
 * a SetAttr(U,ha,a) before IsSet(ha,a) and any occurrence of
 * UnsetAttr(W,ha,a) before IsSet(ha,a) is also before SetAttr(U,ha,a).
 *
 * Intuitively: if there are many occurrences of SetAttr(U,ha,a) and 
 * UnsetAttr(U,ha,a), it must be that SetAttr(U,ha,a) is the last one 
 * before IsSet(ha,a).
 */
restriction IsSet:
"
All ha a #i . IsSet(ha,a)@i ==> 
(
  Ex U #j . SetAttr(U,ha,a)@j & j<i &
  (All W #w . UnsetAttr(W,ha,a)@w & w<i ==> w<j)
)
"

/*
 * An attribute a is unset over ha, written IsUnset(ha,a) if 
 * there exist an UnsetAttr(U,ha,a) before IsSet(ha,a) and any 
 * SetAttr(W,ha,a) before IsSet(ha,a) is also before UnsetAttr(U,ha,a)
 * (this part is the dual of IsSet above)
 * OR
 * there exist no SetAttr(U,ha,a) before IsUnset(ha,a).
 */
restriction IsUnset:
"
All ha a #i . IsUnset(ha,a)@i ==> 
(
  (Ex U #j . UnsetAttr(U,ha,a)@j & j<i &
    (All W #w . SetAttr(W,ha,a)@w & w<i ==> w<j))
  |
  (not Ex U #j . SetAttr(U,ha,a)@j & j<i)
)
"

/*
 * Restricting SO behavior:
 * 
 * If a SO makes ha trusted then the key have been created as 
 * non extractable by a KM.
 */
restriction SO:
"
All W ha #i . SetAttr(W,ha,'trusted')@i
==>
Ex k U #j #w . 
  CreateNEKey(U,ha,k)@j & j<i &
  NewKM(U)@w & w<i
"

/*
 * Restricting KM behavior:
 *
 * KM can only make candidate keys wrap or unwrap.
 */
restriction KM:
"
All U ha k a #i #j #w.  
  NewKM(U)@i &
  CreateNEKey(U,ha,k)@j &
  SetAttr(U,ha,a)@w
==> 
  ( a='wrap' | a='unwrap' )
"

/*
 * This lemma is necessary to make the analysis convergent and efficient. 
 * Tamarin cannot compute the possible sources of terms (because of partial
 * deconstructions) and the lemma tells that unwrapped keys come from previous 
 * API operations such as CreateKey, CreateWWTKey and ImportKey. CreateNEKey 
 * is not possible as it generates non extractable keys.
 */
lemma Unwrap [sources, reuse]:
"
All U k1 ha1 ha2 #i . 
  Unwrap(U,ha1,ha2)@i & 
  IsHandle(ha1,k1)@i
==> 
(
  (Ex W ha #j . CreateKey(W,ha,k1)@j & j<i)
  |
  (Ex W ha #j . CreateWWTKey(W,ha,k1)@j & j<i)
  |
  (Ex W ha #j . ImportKey(W,ha,k1)@j & j<i)
)
"


/* 
 * Sanity:
 *
 * Besides simple sanity lemmas that check that the model really does something
 * (i.e., they check that all the actions are possible, see below) we prove 
 * that the five rules presented in the paper (cf. Section 3.4) are respected.
 */

/*
 * Rule 1: (Sensitive keys). Any sensitive key stored in the HSM should either 
 * be generated with attribute wrap_with_trusted set, or with attribute 
 * extractable unset. 
 *
 * We have two specific actions corresponding to the above cases which set the 
 * appropriate attributes. Notice that we will prove the secrecy of these 
 * particular cases (cf. lemmas SecrecyWWT and SecrecyNE). We also have the
 * case of trusted keys. In particular:
 *
 * - CreateWWTKey: creates a key that has wrap_with_trusted set. We additionally 
 *                 check that the attribute cannot be unset (SanityRule1_1)
 *
 * - CreateNEKey:  creates a key that has extractable unset. We actually 
 *                 check that the attribute can never be set (SanityRule1_2)
 *
 * - Trusted keys: trusted keys should be sensitive. We prove that they are
 *                 are never set as extractable in their whole life-cycle 
 *                 (SanityRule1_3)
 */

lemma SanityRule1_1:
"
All U ha k #i . 
  CreateWWTKey(U,ha,k)@i 
==> (
  SetAttr(U,ha,'wrap_with_trusted')@i &
  not Ex W #j . UnsetAttr(W,ha,'wrap_with_trusted')@j
)
"

lemma SanityRule1_2:
"
All U ha k #i . 
  CreateNEKey(U,ha,k)@i 
==> not Ex W #j . SetAttr(W,ha,'extractable')@j
"

lemma SanityRule1_3:
"
All U ha k #i #j. 
  IsHandle(ha,k)@i & 
  SetAttr(U,ha,'trusted')@j 
==> not Ex W #w . SetAttr(W,ha,'extractable')@w
"

/*
 * Rule 2: (Trusted keys). The SO sets attribute trusted only on special 
 * candidate keys generated by one of the KMs.
 *
 * We check that: 
 *
 * - only the SO can set the attribute trusted (SanityRule2_1)
 *
 * - candidate keys are generated by a KM with extractable unset (SanityRule2_2)
 */
lemma SanityRule2_1:
"
All W ha #i . SetAttr(W,ha,'trusted')@i
==> Ex #j . NewSO(W)@j& j<i
"

lemma SanityRule2_2:
"
All W ha #i . SetAttr(W,ha,'trusted')@i
==> (
  Ex U k #j #w . 
    NewKM(U)@j & j<i & 
    CreateNEKey(U,ha,k)@w & j<w & w<i
)
"

/*
 * Rule 3: (Roles of candidate keys). The candidate keys managed by the KMs 
 * should only admit wrap and unwrap cryptographic operations during their 
 * whole lifetime.
 */
lemma SanityRule3:
"
All U W ha a #i #j #w. 
  SetAttr(W,ha,'trusted')@i & 
  SetAttr(U,ha,a)@j & 
  NewKM(U)@w 
==> ( a = 'wrap' | a = 'unwrap' )
"

/* 
 * Rule 4: (Management of candidate keys). The candidate keys managed by the 
 * KMs should be generated with the extractable attribute unset.
 *
 * We actually check that candidate keys have the extractable attribute unset 
 * during their whole lifetime.
 */

lemma SanityRule4:
"
All W ha #i . SetAttr(W,ha,'trusted')@i
==> (
  (Ex U #j . NewKM(U)@j & j<i) &
  (not Ex U #j . SetAttr(U,ha,'extractable')@j )
)
"

/* 
 * Rule 5: (Freshness of candidate keys). The candidate keys managed by the 
 * KMs should be generated as fresh in the device.
 *
 * This comes implicitly from the definition of rule CreateNEKey that requires
 * Fr(k) in the premise.
 */

/*
 * The following sanity lemmas are important to check that the model really 
 * does something.
 * We check that 
 * - all the actions are possible (the model can be executed). These lemmas 
 *   are of the form exists-trace as they look for one trace satisfying the
 *   required formula;
 * - some intuitive constraints hold (see comments).
 */

lemma SanityUsers:
exists-trace
"
Ex U1 U2 U3 #i1 #i2 #i3.
  NewSO(U1)@i1 &
  NewNU(U2)@i2 &
  NewKM(U3)@i3 
"

/* 
 * Users can only have one role. This is guaranteed by the freshness
 * of U in rules NewNU, NewKM, NewSO
 */
lemma SanityUsersRole:
"
All U1 U2 #i1 #i2 . NewSO(U1)@i1 & NewNU(U2)@i2 ==> not U1 = U2 &
All U1 U2 #i1 #i2 . NewSO(U1)@i1 & NewKM(U2)@i2 ==> not U1 = U2 &
All U1 U2 #i1 #i2 . NewNU(U1)@i1 & NewKM(U2)@i2 ==> not U1 = U2 
"

lemma SanityKeys:
exists-trace
"
Ex U ha1 k1 ha2 k2 ha3 k3 ha4 k4 #i1 #i2 #i3 #i4
  .
  CreateKey   (U,ha1,k1)@i1 &
  CreateWWTKey(U,ha2,k2)@i2 &
  CreateNEKey (U,ha3,k3)@i3 &
  ImportKey   (U,ha4,k4)@i4 
"

lemma SanityAttributesWrap:
exists-trace
"
Ex U ha #i #j.
  SetAttr  (U,ha,'wrap')@i &
  UnsetAttr(U,ha,'wrap')@j 
"

lemma SanityAttributesUnwrap:
exists-trace
"
Ex U ha #i #j.
  SetAttr  (U,ha,'unwrap')@i &
  UnsetAttr(U,ha,'unwrap')@j 
"

lemma SanityAttributesEncrypt:
exists-trace
"
Ex U ha #i #j.
  SetAttr  (U,ha,'encrypt')@i &
  UnsetAttr(U,ha,'encrypt')@j 
"

lemma SanityAttributesDecrypt:
exists-trace
"
Ex U ha #i #j.
  SetAttr  (U,ha,'decrypt')@i &
  UnsetAttr(U,ha,'decrypt')@j 
"

lemma SanityAttributesTrusted:
exists-trace
"
Ex U ha #i #j.
  SetAttr  (U,ha,'trusted')@i &
  UnsetAttr(U,ha,'trusted')@j 
"

lemma SanityAttributesExtractable1:
exists-trace
"
Ex U ha #i.
  UnsetAttr(U,ha,'extractable')@i
"

/* 
 * extractable can be set only by the following rules:
 * - CreateKey
 * - CreateWWTKey
 * - ImportKey
 * - Unwrap
 */
lemma SanityAttributesExtractable2:
"
All U ha #i. SetAttr  (U,ha,'extractable')@i 
==> (
  (Ex k . CreateKey(U,ha,k)@i    ) |
  (Ex k . CreateWWTKey(U,ha,k)@i ) |
  (Ex k . ImportKey(U,ha,k)@i    ) |
  (Ex ha2 . Unwrap(U,ha,ha2)@i   )  
)
"

lemma SanityAttributesWWT1:
exists-trace
"
Ex U ha #i .
  SetAttr  (U,ha,'wrap_with_trusted')@i 
"

/* wrap_with_trusted cannot be unset */
lemma SanityAttributesWWT2:
"
All U ha #i .
  UnsetAttr(U,ha,'wrap_with_trusted')@i 
  ==> F
"

lemma SanityWrap :
exists-trace
"
Ex U ha1 ha2  #i . Wrap(U,ha1,ha2)@i
"

lemma SanityWrapWWT :
exists-trace
"
Ex U W ha1 k1 ha2  #i #j.
  Wrap  (U,ha1,ha2)@i & CreateWWTKey(W,ha1,k1)@j & j<i
"

lemma SanityUnwrap :
exists-trace
"
Ex U ha1 ha2  #i .
  Unwrap(U,ha1,ha2)@i 
"

/*
 * Proof of security: 
 *
 * SecrecyTrusted: trusted keys are never leaked
 * SecrecyNE: non extractable keys are never leaked
 * SecrecyWWT: wrap_with_trusted keys are never leaked
 */


/* 
 * Keys which are generated as with extractable unset are never leaked.
 * This lemma speeds-up consistently the proof of next ones so we prove it
 * as first with the "reuse" label.
 */
lemma SecrecyNE [reuse]:
"
All U ha k #i #j . 
  CreateNEKey(U,ha,k)@i & 
  KU(k)@j 
==> F
"

/*
 * Keys which, at some point, are marked as trusted are never leaked.
 */
lemma SecrecyTrusted:
"
All W ha k #i #j #w. 
  IsHandle(ha,k)@i & 
  SetAttr(W,ha,'trusted')@j & 
  KU(k)@w 
==> F
"

/* 
 * Keys which are generated with wrap_with_trusted set are never leaked.
 */
lemma SecrecyWWT:
"
All U ha k #i #j . 
  CreateWWTKey(U,ha,k)@i & 
  KU(k)@j 
==> F
"

end
\end{lstlisting}

\end{document}